\documentclass[12pt]{article}
\usepackage{amssymb,amsmath,epsfig}

\numberwithin{equation}{section}

\begin{document}

\title{\textbf{Charged Vector Particles
Tunneling From A Pair of Accelerating and Rotating and $5D$ Gauged Super-Gravity Black Holes}}

\author{Wajiha Javed$^1$\thanks{wajiha.javed@ue.edu.pk;wajihajaved84@yahoo.com}, G.
Abbas$^2$ \thanks{abbasg91@yahoo.com;ghulamabbas@iub.edu.pk} and
Riasat Ali$^1$\thanks{riasatyasin@gmail.com} \\
$^1$Division of Science and Technology University\\ of Education
Township Campus, Lahore-54590, Pakistan\\
$^2$Department of Mathematics, The Islamia\\
University of Bahawalpur, Bahawalpur, Pakistan.}

\date{}
\maketitle

\begin{abstract}
The aim of this paper is to study the quantum tunneling process for
charged vector particles through the horizons of more generalized
black holes by using Proca equation. For this purpose, we consider a
pair of charged accelerating and rotating black holes with NUT
parameter and a black hole in $5D$ gauged super-gravity theory,
respectively. Further, we study the tunneling probability and
corresponding Hawking temperature for both black holes by using WKB
approximation. We find that our analysis is independent of the
particles species either background black hole geometries are more
generalized.
\end{abstract}

\section{Introduction}

A black hole (BH) is considered as an object which absorbs all the
matter/energy from the environing area into it due to its intense
gravitational field. General relativity (GR) depicts that a BH
swallows all particles that collides the horizon of the BH. In 1974,
Hawking predicted that a BH behaves like a black body having a
specific temperature, known as \textit{Hawking temperature}, which
allows a BH to emit radiation (called \textit{Hawking radiation})
from its horizon by assuming quantum field hypothesis in the
background of the curved spacetime.

A particle's action of quantum mechanical nature is used in order to
calculate Hawking radiation spectrum from different BHs
\cite{R1,R2}. The analysis of Hawking radiation as a quantum
tunneling phenomenon and accretion onto some particular BHs has attracted the attention of many researcher 
\cite{15a}-\cite{20a}. Various efforts have been carried out to examine this
radiation spectrum from BHs by considering quantum mechanics of
scalar, Dirac, fermion and photon particles etc. Many researchers
\cite{R5}-\cite{R8} have been studied vector particles tunneling to
obtain more information about the Hawking temperature and radiation
spectrum from different BHs. The charged vector particles tunneling
from Kerr-Newman BH \cite{R3} and charged black string \cite{R4} are
important contribution towards the BH physics.

The charged fermions tunneling from Reissner-Nordstr\"{o}m de-Sitter
BH with a global monopole \cite{R9} is studied by using WKB
approximation and Dirac equation to evaluate the tunneling process
for charged particles as well as Hawking temperature. In this paper
the authors have evaluated the tunneling probability and Hawking
temperature for charged fermion tunneling from event horizon. The
tunneling process for Plebanski-Demianski BHs is determined by the
graphical behavior of Hawking temperature of ingoing and outgoing
charged fermion from event horizon \cite{R10}. The Hawking
temperature for charged NUT (Newman-Unti-Tamburino) BH solutions to
the field equations, is considered with rotation and acceleration. A
BH can be studied on the small measurement through quantum field
theory on a curved background \cite{R11}. The tunneling probability
for outgoing particle is ruled by the imaginary part of particle's
action. A large number of attempts \cite{R31}-\cite{R35} have been
made to calculate tunneling of charged and uncharged scalar and
Dirac particles with different BHs configurations. The tunneling of
spin-$\frac{1}{2}$ particles by event horizon of the Rindler
spacetime was explained and Unruh temperature has been calculated
\cite{R12}. Kraus and Wilczek \cite{R14, R15} projected a
semi-classical process to analyze Hawking radiation as a tunneling
event. This process contains the calculation for the phenomenon of
s-wave emission across event horizon. In \cite{R16}, it has been
shown that the Hawking radiation from rotating wormhole may emit all
types of particles.

This paper deals with the study of the Hawking radiation process of
charged vector particles from the horizons of a pair of accelerating
and rotating BHs and a BH in $5D$ gauged super-gravity.

Vector particles (spin-1 bosons) such as $Z$ (uncharged) and
$W^{\pm}$ (charged) bosons are of very importance in Standard Model.
In the background of BHs geometries, the behavior of the bosons can
be determined by using Proca equation. First, we formulate the field
equations of charged $W^{\pm}$-bosons by using Lagrangian of the
\textmd{Glashow}-\textmd{Weinberg}-\textmd{Salam model} \cite{R21}.
Then we shall investigate particle emission process by using the
Hamilton-Jacobi definition and WKB approximation to the derived
equation for charged case in the considered BHs geometries. By
putting the determinant (of coefficient matrix) equals to zero, we
can solve for radial function. Consequently, we compute the
tunneling rate of the charged vector particles from the horizons of
BHs and find the corresponding Hawking temperature values in both
cases.

The paper is planned follows: We discuss in the section \textbf{2},
the tunneling rate and Hawking temperature for charged accelerating
and rotating BH solutions with NUT parameter. Section \textbf{3} is
devoted to investigate the charged vector particles tunneling and
Hawking temperature for BH in 5D gauged super-gravity spacetime, by
investigating the $W^{\pm}$ bosons observation. Section \textbf{4}
provides the summary of the results for both cases.

\section{Accelerating and Rotating Black Holes with NUT Parameter}

In universal, the NUT parameter is affiliated with the
gravitomagnetic monopole, related to the bending properties of the
environing spacetime due to the fundamental mass, its accurate
physical significance could not be determined. The generalization
for multi dimensional Kerr-NUT de-Sitter spacetime \cite{R22, R23}
and its physical implication \cite{R24} is also investigated. As a
BH, the dominance on the NUT parameter the revolution parameter
departs the spacetime free on bending singularities and the agreeing
result is appointed as NUT alike result. If the revolution parameter
commands the NUT parameter, the result is Kerr-like and a closed
chain bending singularity forms. The behavior of this form of the
singularity structure is independent of the existence on the
cosmology constant.

There are lots of BHs which comprise of the NUT parameter and lots
of investigation have been made to examine their physical effects in
the space of colliding waves. Accurate significance of the NUT
parameter exists, when a motionless Schwarzschild mass is absorbed
in a stationary source and allows electromagnetic universe
\cite{R25}. The NUT parameter is referred to the bend of the
electromagnetic universe leaving out the fundamental Schwarzschild
mass. In the absence of electromagnetic field, it reduces to the
bend of the vacuum spacetime \cite{R26}. The bend of the surrounding
space pair with the mass of reference yields NUT parameter.

The line element for accelerating and rotating BHs with NUT
parameter is defined as \cite{R27}
\begin{eqnarray}\nonumber
ds^{2}&=&-\frac{1}{\Omega^{2}}\Big[\frac{Q}{\rho^{2}}\Big(dt-
(a\sin^{2}\theta+4l\sin^{2}\frac{\theta}{2})d\phi\Big)^{2}
-\frac{\rho^{2}}{Q}dr^{2}\nonumber\\&& \frac{\tilde{P}}{\rho^{2}}
\Big(a dt-(r^{2}+(a+l)^{2})d\phi\Big)^{2}
-\frac{\rho^{2}}{\tilde{P}}\sin^{2}\theta d\theta^{2}\Big],\label{1}
\end{eqnarray}
where
\begin{eqnarray}
\Omega&=&1-\frac{\alpha}{\omega}(l+a\cos\theta)r
,~~~\rho^{2}=r^{2}+(l+a\cos\theta)^{2},\nonumber\\
Q&=&\left[(\omega^{2}\tilde{k}+\tilde{e}^{2}+\tilde{g}^{2})(1+2\alpha
l\frac{r}{\omega})-2Mr+\frac{\omega^{2}\tilde{k}r^{2}}{a^{2}-l^{2}}\right]\nonumber\\
&\times&\left[1+\alpha\frac{a-l}{\omega}r\right]\left[1-\alpha\frac{a+l}{\omega}r\right],\nonumber\\
\tilde{P}&=&\sin^{2}\theta(1-a_{3}\cos\theta-a_{4}\cos^{2}\theta)=P\sin^{2}\theta,\nonumber\\
a_{3}&=&2M\frac{\alpha a}{\omega}-4\frac{al\alpha^{2}}{\omega^{2}}(\omega^{2}\tilde{k}+\tilde{e}^{2}+\tilde{g}^{2}),\nonumber\\
a_{4}&=&-\frac{\alpha^{2}a^{2}}{\omega^{2}}(\omega^{2}\tilde{k}+\tilde{e}^{2}+\tilde{g}^{2}).\nonumber
\end{eqnarray}
Here, $M$ denotes the mass of pairs of BHs, $e$ and $g$ indicate the
electric and magnetic charges, respectively, while $l$ is a NUT
parameter of BH, $\alpha$ and $\omega$ indicate acceleration and
rotation of sources, respectively. Also, $a$ is the Kerr-like
rotation parameter and $\tilde{k}$ is given by
\begin{equation}
\left(\frac{\omega^{2}}{a^{2}-l^{2}}+3\alpha^{2}l^{2}\right)\tilde{k}=1+2\frac{\alpha
l}
{\omega}M-3\frac{\alpha^{2}l^{2}}{\omega^{2}}(\tilde{e}^{2}+\tilde{g}^{2}).\nonumber
\end{equation}
Here, $\alpha$, $\omega$, $M, \tilde{e}, \tilde{g}$ and $\tilde{k}$
are arbitrary real parameters. We would like to mention that
$\omega$ depends on NUT parameter $l$ and Kerr-like rotation
parameter $a$. The $\alpha$ twisting property of BHs is proportional
to the rotation $\omega$. Also, $\omega$ depends on rotation
parameters $l$ and $a$. The parameters $\alpha$, $\omega$, $M,
\tilde{e}, \tilde{g}$ and $\tilde{k}$ vary independently. If
$\alpha$ is equal to zero, then metric in Eq.(\ref{1}) leads to the
Kerr-Newman-NUT solution. If $l=0$, then metric in Eq.(\ref{1})
gives the couple of charged and rotating BHs. In this case, if
$\tilde{e}$ and $\tilde{g}$ are equal to zero, we have a
Schwarzschild BH and if $l$ and $a$ are equal to zero it leads to
C-metric.

The metric (\ref{1}) can be rewritten as
\begin{equation}
ds^{2}=-f(r,\theta)dt^{2}+\frac{dr^{2}}{g(r,\theta)}
+\Sigma(r,\theta)d\theta^{2}+k(r,\theta)d\phi^{2}
-2{H}(r,\theta)dtd\phi,
\end{equation}
where $f(r,\theta)$, $g(r,\theta)$, $\Sigma(r,\theta)$,
$K(r,\theta)$ and $H(r,\theta)$ are given by the following
equations:
\begin{eqnarray*}
f(r,\theta)&=&\frac{Q-Pa^2\sin^2\theta}{\rho^2\Omega^2},\quad
g(r,\theta)=\frac{Q\Omega^2}{\rho^2},\quad\Sigma=(r,\theta)=\frac{\rho^2}{\Omega^2P},\\
k(r,\theta)&=&\frac{1}{\Omega^2\rho^2}\Big(sin^2\theta
P(r^2+(a+l)^2)^2-Q(a\sin^2\theta+4l\sin^2\frac{\theta}{2})^2\Big),\\
H(r,\theta)&=&\frac{1}{\Omega^2\rho^2}\Big(sin^2\theta
Pa(r^2+(a+l)^2)-Q(a\sin^2\theta+4l\sin^2\frac{\theta}{2})\Big).
\end{eqnarray*}
The electromagnetic potential for these BHs is given
\begin{eqnarray}
A=&&\frac{1}{a(r^2+(l+acos\theta)^2)}\Big[-\tilde{e}r\Big(adt-d\phi((l+a)-
(l^2+a^2\cos^2\theta+2al\cos\theta))\Big)\nonumber\\&&
-\tilde{g}\Big(l+a\cos\theta\Big(a
dt-d\phi\Big(r^2+(l+a)^2)\Big)\Big].
\end{eqnarray}
The event horizons are obtained for
$g(r,\theta)=\frac{Q\Omega^2}{\rho^2}=0$, which implies that
$\Omega\neq0$, so $Q=0$, which yields the following real roots of
$r$, i.e.,
\begin{eqnarray}
&&r_{\alpha1}=\frac{\omega}{\alpha(a+l)},\quad
r_{\alpha2}=\frac{-\omega}{\alpha(a-l)}\quad
r_{\pm}=\frac{a^2-l^2}{\omega^2\tilde{k}}\Big[-\Big((\omega^2\tilde{k}+\tilde{e}^2+\tilde{g}^2)\frac{\alpha
l}{\omega}-M\Big)\nonumber\\&&\pm
\sqrt{\Big((\omega^2\tilde{k}+\tilde{e}^2+\tilde{g}^2)\frac{\alpha
l}{\omega}-M\Big)^2-(\omega^2\tilde{k}+\tilde{e}^2+\tilde{g}^2)\frac{\omega^2\tilde{k}}{\alpha^2
-l^2}} \Big],
\end{eqnarray}
where $r_{\alpha1}$ and $r_{\alpha2}$ are acceleration horizons and
$r_\pm$ represent the outer and inner horizons, respectively such
that
\begin{equation*}{\Big((\omega^2\tilde{k}+\tilde{e}^2+\tilde{g}^2)\frac{\alpha
l}{\omega}-M\Big)^2-(\omega^2\tilde{k}+\tilde{e}^2+\tilde{g}^2)\frac{\omega^2\tilde{k}}{a^2
-l^2}}>0.
\end{equation*}
The angular velocity at BH outer (event)
horizon is defined by
\begin{equation}
{\check{\Omega}}= \frac{a}{r^2_++(a+l)^2}.
\end{equation}

In order to investigate the tunneling spectrum for charged vector
particles through the BH horizon, we will consider Proca equation
with electromagnetic effects. In a curved spacetime with
electromagnetic field, the motion of massive spin-1 charged vector
fields is depicted by the given Proca equation by using the
Lagrangian of the W-bosons of Glashow-Weinberg-Salam model \cite{R6}
\begin{equation}
\frac{1}{\sqrt{-\textbf{g}}}\partial_{\mu}(\sqrt{-g}\psi^{\nu\mu})+
\frac{m^{2}}{h^{2}}\psi^{\nu}+\frac{i}{h}e A_{\mu}\psi^{\nu\mu}+
\frac{i}{h}eF^{\nu\mu}\psi_{\mu}=0,\label{2}
\end{equation}
where $\textbf{g}$ is determinant of coefficients matrix, $m$ is
particles mass and $\psi^{\mu\nu}$ is anti-symmetric tensor, i.e.,
\begin{equation}
\psi_{\nu\mu}=\partial_{\nu}\psi_{\mu}-\partial_{\mu}\psi_{\nu}+
\frac{i}{h}e A_{\nu}\psi_{\mu}-\frac{i}{h}e A_{\mu}\psi_{\nu}~~
\textmd{and}~~
F^{\mu\nu}=\nabla^{\mu}A^{\nu}-\nabla^{\nu}A^{\mu}.\nonumber
\end{equation}
Here, $A_\mu$ is considered as the electromagnetic potential of the
BH, $e$ denotes the charge of the W-bosons and $\nabla_{\mu}$ is
geometrically covariant derivative. Since the equation of motion for
the $W^+$ and $W^-$ bosons is similar, the tunneling processes
should be similar too. For simplification, here we will consider the
$W^+$ boson case, the results of this case can be extended to $W^-$
bosons due to the digitalization of the line element. For $W^+$
field, the values of the components of $\psi^{\mu}$ and
$\psi^{\nu\mu}$ are obtained as follows
\begin{eqnarray}
&&\psi^{0}=\frac{-k\psi_{0}-H\psi_{3}}{f k+H^{2}},~~~ \psi^{1}=g
\psi_{1},~~~ \psi^{2}=\Sigma^{-1}
\psi_{2},~~~\psi^{3}=\frac{-H\psi_{0}+f\psi_{3}}{f k+H^{2}},\nonumber\\
&&\psi^{01}=\frac{-kg\psi_{01}-Hg\psi_{13}}{f
k+H^{2}},~~~\psi^{02}=\frac{-k\psi_{02}-H\psi_{23}}{\Sigma(f
k+-H^{2})},~~~\psi^{03}=\frac{-\psi_{03}}{f k+H^{2}},\nonumber\\
&&\psi^{12}=g\Sigma^{-1}\psi_{12},~~~\psi^{13}=\frac{g(f\psi_{13}-H\psi_{01})}{f
k+H^{2}},~~~\psi^{23}=\frac{g\psi_{23}-H\psi_{02}}{\Sigma(f
k+H^{2})}.\nonumber
\end{eqnarray}
The electromagnetic vector potential for this BH is given by
\cite{R28}
\begin{eqnarray}
A&=&\frac{1}{a[r^{2}+(l+a\cos\theta)^{2}]}[-\tilde{e}r[a
dt-d\phi(l+a)^{2}-(l^{2}+a^{2}\cos^{2}\theta\nonumber\\
&+&2la\cos\theta)]-\tilde{g}(l+a\cos\theta)[a
dt-d\phi{r^{2}+(l+a)^{2}}]].
\end{eqnarray}
Using, the WKB approximation \cite{R29}, i.e.,
\begin{equation}
\psi_{\nu}=c_{\nu}\exp[\frac{i}{\hbar}S_{0}(t,r,\theta,\phi)+
\Sigma \hbar^{n}S_{n}(t,r,\theta,\phi)],
\end{equation}
to the Proca Eq.(\ref{2}) and neglecting the terms for
$n=1,2,3,4,...$, we obtain the following set of equations
\begin{eqnarray}
&kg&[c_{1}(\partial_{1}S_{0})((\partial_{1}S_{0})+e A_{0})
-c_{0}(\partial_{1}S_{0})^{2}]-H g[c_{3}(\partial_{1}S_{0})^{2}\nonumber\\
&-&c_{1}(\partial_{1}S_{0})((\partial_{3}S_{0})
+eA_{3})]+\frac{k}{\Sigma}\left[c_{2}(\partial_{2}S_{0})((\partial_{0}S_{0})
+eA_{3})-c_{0}(\partial_{2}S_{0})^{2}\right]\nonumber\\
&-&\frac{H}{\Sigma}\left[c_{3}(\partial_{2}S_{0})^{2}-c_{2}
(\partial_{2}S_{0})((\partial_{3}S_{0})+eA_{3})\right]
+[c_{3}(\partial_{3}S_{0})((\partial_{0}S_{0})+eA_{0})\nonumber\\
&-&c_{0}(\partial_{3}S_{0})((\partial_{3}S_{0})+eA_{3})]+eA_{3}kg[c_{1}
((\partial_{0}S_{0})+eA_{0})-c_{0}(\partial_{1}S_{0})]\nonumber\\
&-&m^{2}kc_{0}-m^{2}Hc_{3}-eA_{3}Hg[c_{3}
(\partial_{1}S_{0})-c_{1}((\partial_{3}S_{0})+eA_{3})]=0,\label{3}\\
&kg&[c_{1}(\partial_{0}S_{0})((\partial_{0}S_{0})+eA_{0}-c_{0}(\partial_{1}S_{0})
(\partial_{0}S_{0})]-Hg[c_{3}(\partial_{1}S_{0})(\partial_{0}S_{0})\nonumber\\
&-&c_{1}(\partial_{0}S_{0})((\partial_{3}S_{0})+eA_{3})]+\frac{g(fk+H^{2})}{\Sigma}\left[c_{2}
(\partial_{1}S_{0})(\partial_{2}S_{0})-c_{1}(\partial_{2}S_{0 })^{2}\right]\nonumber\\
&+&gf[c_{3}(\partial_{1}S_{0})(\partial_{3}S_{0})-c_{1}(\partial_{3}S_{0})
((\partial_{3}S_{0})+eA_{3})]+gH[c_{1}(\partial_{3}S_{0})((\partial_{0}S_{0})\nonumber\\
&-&eA_{0})-c_{0}(\partial_{1}S_{0})(\partial_{3}S_{0})]+eA_{0}kg[c_{1}((\partial_{0}S_{0})
-eA_{0})-c_{0}(\partial_{1}S_{0})]\nonumber\\
&-&m^{2}gc_{1}(fk-H)-eA_{0}Hg[c_{3}(\partial_{1}S_{0})-c_{1}((\partial_{3}S_{0})-eA_{3})]
\nonumber\\&+&eA_{3}gf[c_{3}(\partial_{1}S_{0})-c_{1}((\partial_{3}S_{0})+eA_{3})]+eA_{3}
H[c_{1}((\partial_{0}S_{0})+eA_{0})\nonumber\\&-&c_{0}(\partial_{1}S_{0})]=0\label{4}\\
&&{\frac{k}{\Sigma}}\left[c_{2}(\partial_{2}S_{0})^{2}-c_{0}(\partial_{2}S_{0})(\partial_{0}S_{0})
+eA_{0}(\partial_{0}S_{0})c_{2}\right]-\frac{H}{\Sigma}[c_{3}(\partial_{1}S_{0})(\partial_{2}S_{0})\nonumber\\
&-&c_{2}(\partial_{1}S_{0})(\partial_{3}S_{0})]-\frac{g}{\Sigma}[c_{2}(\partial_{1}S_{0})^{2}
-c_{1}(\partial_{1}S_{0})(\partial_{2}S_{0})](fk-H)\nonumber\\
&+&\frac{f}{\Sigma}[c_{3}(\partial_{2}S_{0})(\partial_{3}S_{0})-c_{2}(\partial_{3}S_{0})^{2}
-eA_{3}c_{2(\partial_{3}S_{0})}]+\frac{H}{\Sigma}[(\partial_{3}S_{0})(\partial_{0}S_{0})c_{2}
\nonumber\\&-&c_{0}(\partial_{2}S_{0})(\partial_{3}S_{0})+c_{2}eA_{0}(\partial_{2}S_{0})(\partial_{3}S_{0})]
-m^{2}c_{2}{\Sigma}^{-1}(fk-H)\nonumber\\&+&eA_{0}\frac{k}{\Sigma}[c_{2}(\partial_{0}S_{0})-
c_{0}(\partial_{2}S_{0})+eA_{0}c_{2}]-eA_{0}\frac{H}{\Sigma}[c_{3}(\partial_{2}S_{0})-c_{2}
(\partial_{3}S_{0})\nonumber\\&-&c_{2}eA_{3}]+eA_{3}\frac{f}{\Sigma}[c_{3}(\partial_{2}S_{0})-c_{2}(\partial_{3}S_{0})
-eA_{3}c_{2}]+eA_{3}\frac{H}{\Sigma}[c_{2}(\partial_{0}S_{0})\nonumber\\
&-&c_{0}(\partial_{2}S_{0})+eA_{0}c_{2}]=0\label{5}
\end{eqnarray}
\begin{eqnarray}
&&[c_{3}(\partial_{0}S_{0})^{2}-c_{0}(\partial_{3}S_{0})(\partial_{0}S_{0})+
eA_{0}c_{3}(\partial_{0}S_{0})-eA_{3}c_{0}(\partial_{0}S_{0})]\nonumber\\
&+&g-H[c_{1}(\partial_{0}S_{0})(\partial_{1}S_{0})
-c_{0}(\partial_{1}S_{0})^{2}+eA_{0}c_{1}(\partial_{1}S_{0})]-fg[c_{3}(\partial_{2}S_{0})^{2}\nonumber\\
&-&c_{1}(\partial_{1}S_{0})(\partial_{3}S_{0})-eA_{3}c_{1}(\partial_{0}S_{0})]-
\frac{H}{\Sigma}[c_{2}(\partial_{0}S_{0})(\partial_{2}S_{0})-c_{0}(\partial_{2}S_{0})^{2}\nonumber\\
&+&eA_{0}c_{2}(\partial_{2}S_{0})]-fg[c_{3}(\partial_{1}S_{0})^{2}-c_{1}
(\partial_{1}S_{0})(\partial_{3}S_{0})-eA_{3}c_{1}(\partial_{0}S_{0})]\nonumber\\
&-&\frac{H}{\Sigma}[c_{2}(\partial_{0}S_{0})(\partial_{2}S_{0})-
c_{0}(\partial_{2}S_{0})^{2}+eA_{0}c_{2}(\partial_{2}S_{0})]
-\frac{f}{\Sigma}[c_{3}(\partial_{2}S_{0})^{2}\nonumber\\
&-&c_{2}(\partial_{2}S_{0})(\partial_{3}S_{0})-eA_{3}c_{2}(\partial_{2}S_{0})]
+eA_{0}[c_{3}(\partial_{0}S_{0})-c_{0}(\partial_{3}S_{0})\nonumber\\
&+&eA_{0}c_{3}-eA_{3}c_{0}+m^{2}[Hc_{0}+c_{3}f]=0.\label{6}
\end{eqnarray}
Using separation of variables technique, we can choose
\begin{equation}
S_{0}=-(E-j\check{\Omega})t+W(r)+N\phi+\Theta(\theta),
\end{equation}
where $E$ and $j$ represent particle's energy and angular momentum,
respectively. From the Eqs.(\ref{3})-(\ref{6}), we can obtain a
matrix equation
\begin{equation*}
G(c_{0},c_{1},c_{2},c_{3})^{T}=0,
\end{equation*}
which implies a $4\times4$ matrix labeled as "$G$", whose components
are given as follows:
\begin{eqnarray}
G_{11}&=&-\dot{W}^{2}kg-\frac{kN^{2}}{\Sigma}-\dot{\Theta}^{2}-\dot{\Theta}eA_{3}
-m^{2}k-eA_{3}kg\dot{W},\nonumber\\
G_{12}&=&-\dot{W}kg(E-j\check{\Omega})+kg\dot{W}eA_{0}+Hg\dot{W}\dot{\Theta}
+Hg\dot{W}eA_{3}\nonumber\\
&-&eA_{3}kg(E-j\check{\Omega})+kge^{2}A_{3}A_{0}+eA_{3}Hg\dot{\Theta}+Hge^{2}A_{3},\nonumber\\
G_{13}&=&-\frac{k}{\Sigma}(E-j\check{\Omega})N+\frac{k}{\Sigma}NeA_{3}+
\frac{H}{\Sigma}\dot{\Theta}N+\frac{H}{\Sigma}NeA_{3},\nonumber\\
G_{14}&=&-\dot{W}^{2}Hg-\frac{H}{\Sigma}N^{2}-\dot{\Theta}^{2}
(E-j\check{\Omega})+\dot{\Theta}eA_{0}-m^{2}H-eA_{3}gH\dot{W},\nonumber\\
G_{21}&=&kg(E-j\check{\Omega})\dot{W}-gH\dot{W}\dot{\Theta}-
eA_{0}kg\dot{W}-eA_{3}H\dot{W},\nonumber\\
G_{22}&=&-kg(E-j\check{\Omega})(-(E-j\check{\Omega})+eA_{0}-gH
(E-j\check{\Omega})(\dot{\Theta}+eA_{3})\nonumber\\
&+&\frac{g}{\Sigma}N^{2}(fk-H)-gf\dot{\Theta}(\dot{\Theta}+eA_{3}))
+gH\dot{\Theta}(-(E-j\check{\Omega})\nonumber\\
&-&eA_{0})+eA_{0}kg(-(E-j\check{\Omega})+eA_{0}gH(\dot{\Theta}-
eA_{3})-eA_{3}gf(\dot{\Theta}+eA_{3})\nonumber
\end{eqnarray}
\begin{eqnarray}
 &&eA_{3}H(-(E-j\check{\Omega})-m^{2}g(fk-H),~~~G_{23}=
\frac{g}{\Sigma}\dot{W}N(fk-H^{2}),\nonumber\\
G_{24}&=&Hg(E-j\check{\Omega})\dot{W}+gf\dot{W}\dot{\Theta}-
eA_{0}gH\dot{W}+eA_{3}gf\dot{W},\nonumber\\
G_{31}&=&\frac{k}{\Sigma}(E-j{\check{\Omega}})N-\frac{H}
{\Sigma}N\dot{\Theta}-eA_{3}HN,~~~G_{32}=\frac{g}{\Sigma}\dot{W}N(E-j{\check{\Omega}}),\nonumber\\
G_{33}&=&-\frac{k}{\Sigma}[(E-j{\check{\Omega}})^{2}-eA_{0}
(E-j{\check{\Omega}})]+\frac{H}{\Sigma}\dot{W}\dot{\Theta}\nonumber\\&-&\frac{g}{\Sigma}
[\dot{W}^{2}(fk+H^{2})]-\frac{f}{\Sigma}[(\dot{\Theta})^{2}+eA_{3}(\dot{\Theta})]+
\frac{H}{\Sigma}[\dot{\Theta}((E-j{\check{\Omega}})
+eA_{0}N)]\nonumber\\&-&m^{2}\Sigma^{-1}(fk+H^{2})-eA_{0}\frac{k}{\Sigma}
[(E-j{\check{\Omega}})-eA_{0}]+eA_{0}\frac{H}{\Sigma}[\dot{\Theta}+eA_{3}]\nonumber\\&-&
eA_{3}\frac{f}{\Sigma}[\dot{\Theta}+eA_{3}],\nonumber\\
G_{34}&=&-\frac{H}{\Sigma}N\dot{W}+\frac{f}{\Sigma}N\dot{\Theta}-eA_{0}N\frac{H}{\Sigma}+
eA_{3}+N\frac{f}{\Sigma},\nonumber\\
G_{41}&=&(E-j{\check{\Omega}})\dot{\Theta}+(E-j{\check{\Omega}})eA_{3}+gH\dot{W}^{2}+N^{2}\frac{H}{\Sigma}
+m^{2}H+(E-j{\check{\Omega}})eA_{0}\nonumber\\&-&e^{2}A_{0}A_{3},~~~G_{42}=gH\dot{W}
(E-j{\check{\Omega}})-gHeA_{0}\dot{W}+fg\dot{W}\dot{\Theta}\nonumber\\
&-&fgeA_{3}(E-j{\check{\Omega}}),\nonumber\\
G_{43}&=&\frac{H}{\Sigma}(E-j{\check{\Omega}})N-\frac{H}{\Sigma}NeA_{0}
+\frac{f}{\Sigma}N\dot{\Theta}+N eA_{3},\nonumber\\
G_{44}&=&(E-j{\check{\Omega}})^{2}-(E-j{\check{\Omega}})eA_{0}
-fg\dot{W}^{2}-\frac{f}{\Sigma}N-m^{2}f\nonumber\\
&-&eA_{0}[(E-j{\check{\Omega}})-eA_{0}],\nonumber
\end{eqnarray}
where $\dot{W}=\partial_{r}S_{0}$,
$\dot{\Theta}=\partial_{\theta}S_{0}$ and $N=\partial_{\phi}S_{0}$.
For the non-trivial solution, the absolute value $\textbf{G}$ is
equals to zero, and solving the resultant equation for the radial
part so that one can get the following integral
\begin{equation}\label{a1}
ImW^{\pm}=\pm \int\sqrt{\frac{(E-eA_{0}-j{\check{\Omega}})^{2}+X}{f(r)g(r)}}dr
\end{equation}
where $+$ and $-$ represent the radial function of outgoing and
incoming particles, respectively, while the function $X$ can be
defined as
$X=-\Sigma^{-1}fN-m^{2}f-Hg(E-j{\check{\Omega}})-gf\dot{\Theta}+
eA_{0}gH-eA_{3}gf, {\check{\Omega}}$ is the angular velocity on the event
horizon.

Expanding the functions $f(r)$ and $g(r)$ in Taylor's series near horizon, we get
\begin{equation}\label{a2}
f(r_+)\approx f'(r_+)(r-r_+),\quad\quad g(r_+)\approx g'(r_+)(r-r_+).
\end{equation}
Using above expressions in Eq.(\ref{a1}), one can see that resulting equation
has we poles at $r=r_+$. For the calculation of Hawking temperature
by using tunneling method, it is required to regularize the singularity by a
specific complex contour to bypass the pole. For our standard co-ordinates
of BH metric, the tunneling of out going particles can be obtained by taking
an infinitesimal halfcircle below the pole $r=r_+$, while for the ingoing
particle such contour is taken below above the pole.
Further, in order to calculate the semiclassical tunneling probability,
it is required that resulting wave equation must be multiplied by its complex conjugate.
In this way, the part of trajectory that starts from outside of the BH and
continues to the observer, will not contribute to the calculation of the final
tunneling probability and can be ignored because it will be completely real.
Therefore, the only part trajectory that contributes
to the tunneling probability is the contour around the BH horizon.

Hence using Eq.(\ref{a1}) and Eq.(\ref{a2}), and integrating the resulting equation around the pole, we get
\begin{equation}
ImW^{\pm}
=\pm i\pi\frac{E-eA_{0}-j{\check{\Omega}}}{2\kappa(r_{+})},
\end{equation}
and the surface gravity is \cite{R26}
\begin{equation}
\kappa(r_{+})=\left[\frac{[\frac{\alpha
l}{\omega}(\omega^{2}\tilde{k}+\tilde{e}^{2}+\tilde{g}^{2})
-M+\frac{\omega^{2}\tilde{k}}{a^{2}-l^{2}}r_{+}]}{[r^{2}_{+}+(a+l)^{2}]}
\times[1+\frac{\alpha(a-l)}{\omega}r_{+}]\times[1-\frac{\alpha(a+l)}{\omega}r_{+}]\right].
\nonumber
\end{equation}
The tunneling probability for charged vector particles is given by
\begin{eqnarray}\nonumber
\Gamma=&&\frac{Prob{[emission]}}{Prob{[absorption]}}=
\frac{\textmd{exp}[-2(ImW^++Im\Theta)]}{\textmd{exp}[-2(ImW^--Im\Theta)]}
={\textmd{exp}[-4ImW^+]}\\\nonumber =&&\exp\left[-2\pi
\frac{E-eA_{0}-j{\check{\Omega}}}{[\frac{[\frac{\alpha
l}{\omega}(\omega^{2}\tilde{k}+\tilde{e}^{2}+\tilde{g}^{2})
-M+\frac{\omega^{2}\tilde{k}}{a^{2}-l^{2}}r_{+}]}{[r^{2}_{+}+(a+l)^{2}]}
\times[1+\frac{\alpha(a-l)}{\omega}r_{+}]\times[1-\frac{\alpha(a+l)}{\omega}r_{+}]]}\right].
\end{eqnarray}
Now, finally we can calculate the Hawking temperature by comparing the above result with the Boltzmann formula
 $\Gamma_B= e^{-(E-e A_0-j\check{\Omega})/T_H}$, to get
\begin{equation}\label{aTH}
T_{H}=\Big[\frac{\Big[{[\frac{\alpha
l}{\omega}(\omega^{2}\tilde{k}+\tilde{e}^{2}+\tilde{g}^{2})
-M+\frac{\omega^{2}\tilde{k}}{a^{2}-l^{2}}r_{+}]}
\times[1+\frac{\alpha(a-l)}{\omega}r_{+}][1-\frac{\alpha(a+l)}{\omega}r_{+}]
\Big]}{2\pi[r^{2}_{+}+(a+l)^{2}]}\Big].
\end{equation}
The Hawking temperature depend on $A_{0}$ vector potential, $E$
energy, ${\check{\Omega}}$ angular momentum, $M$ is a mass a pair of
BHs, $e$ and $g$ are electric and magnetic charges respectively, $a$
is the rotation of a BH, $l$ is a NUT parameter, $\alpha$ represents
acceleration of the sources and $\omega$ rotation of the sources.

We would like to mention that Hawking temperature of charged vector
particles given in Eq.(\ref{aTH}) is same as the Hawking temperature
of fermion particles in Eq.(4.20) of \cite{R26}. Thus the Hawking
temperature is independent of the particle species.

\section{Black Holes in 5D Gauged Super-gravity}

The gauged theory is stated as a super-gravity theory in which the
gravitino, the superpartner of the graviton is charged under some
internal gauge group. However, the gauged super-gravity is more
significant as compared to the ungauged case, because this theory
has a negative cosmological constant, so it is defined on an anti-de
Sitter space. Here, for the discussion of charged vector particles
tunneling spectrum form a BH in $5D$ gauged super-gravity, we
evaluate the tunneling probability of particles and the
corresponding Hawking temperature at BH horizon. Such BH solutions
occur in $D=5~~N=8$ gauged super-gravity (symmetry) \cite{R30}.
Firstly, this solution was formulated in \cite{R29a} as a particular
case (STU-model) of solutions of $D=5 N=2$ gauged super-gravity
equations of motion. The metric for BH in $5D$ gauged super-gravity
is \cite{R30}
\begin{equation}
ds^{2}=-\left(H_{1}H_{2}H_{3}\right)^{-\frac{2}{3}}fdt^{2}
+\left(H_{1}H_{2}H_{3}\right)^{\frac{1}{3}}
\left(f^{-1}dr^{2}+r^{2}d\Omega^{2}_{3,k}\right),
\end{equation}
where
\begin{equation*}
f=k-\frac{\mu}{r^{2}}+g^{2}r^{2}H_{1}H_{2}H_{3},~~~H_{i}=1+\frac{q_{i}}{r^{2}},~~(\textmd{for}~
i=1,2,3)
\end{equation*}
and $d\Omega^{2}_{3,k}$ is the metric on $S^{3}$ with unit radius if
$k=1$, or the metric on $\textbf{R}^3$ if $k=0$, here $\mu$ is the
non-extremality parameter \cite{R29a}, which is related to ADM mass,
$g=1/L$ is the inverse radius of $AdS_5$ related to the cosmological
constant $\Lambda=-6g^2=-6/L^2$, and $q_{i}$ are charges entering
the metric. The three gauge field potentials $A^i_\mu$ from the
solution of equation of motion are of the form
\begin{equation*}
A_{0}^{i}=\frac{\tilde{q_{i}}}{r^{2}+q_{i}}~~(\textmd{for}~~
i=1,2,3)
\end{equation*}
where $\tilde{q_{i}}$ are physical charges which are conserved and
Gauss law is applicable to such charges.

The line element can be rewritten as
\begin{equation}\label{5D}
ds^{2}=-\tilde{A}({r})dt^{2}+\tilde{B}^{-1}({r})dr^{2}
+\tilde{C}({r})d\theta^{2}+\tilde{D}({r})d\phi^{2}+
\tilde{E}({r})d\zeta^{2}.
\end{equation}
where
\begin{eqnarray*}
&& \tilde{A}(r)=f(H_1 H_2  H_3)^{-\frac{2}{3}}\quad\quad
 \tilde{B}^{-1}(r)=f^{-1}(H_1 H_2  H_3)^{{\frac{1}{3}}}\\
  &&\tilde{C}(r)=r^2(H_1 H_2  H_3)^{{\frac{1}{3}}}\quad\quad
 \tilde{D}(r)=r^2\sin^2\theta(H_1 H_2  H_3)^{{\frac{1}{3}}}\\
 && \tilde{E}(r)=r^2 \sin^2\theta \sin^2\phi (H_1 H_2  H_3)^{{\frac{1}{3}}}
\end{eqnarray*}

The horizons of metric (\ref{5D}) can be determined when $f(r)=0$.
For this purpose we follow \cite{R30} and assume that $g^2=1$ (by
the choice of units as in \cite{R30}). Hence, in this case the outer
horizon is located at
\begin{eqnarray*}
r_{+}&=&\sqrt{\frac{\sqrt{(1+q_i)^2+4\mu}-(1+q_i)}{2}},\\~~&\textmd{for}&~~\sqrt{(1+q_i)^2+4\mu}>(1+q_i)\quad\textmd{and
}~~ i=1,2,3.
\end{eqnarray*}

In Proca Eq.(\ref{2}) the components of $\psi^{\nu}$ and
$\psi^{\mu\nu}$ are given by
\begin{eqnarray}
\psi^{0}&=&-\tilde{A}^{-1}\psi_{0},~~\psi^{1}=\tilde{B}\psi_{1}
,~~\psi^{2}=\tilde{C}^{-1}\psi_{2},~~
\psi^{3}=\tilde{D}^{-1}\psi_{3},~~\psi^{4}=\tilde{E}^{-1}\psi_{4},\nonumber\\
\psi^{o1}&=&-\tilde{B}\tilde{A}^{-1}\psi_{01},~~\psi^{02}=-(\tilde{A}
\tilde{C})^{-1}\psi_{02},~~\psi^{03}=-(\tilde{A}\tilde{D})^{-1}\psi_{03}\nonumber\\
\psi^{04}&=&-(\tilde{A}\tilde{E})^{-1}\psi_{04},~~
\psi^{12}=\tilde{B}\tilde{C}^{-1}\psi_{12},~~\psi^{13}=\tilde{B}\tilde{D}^{-1}\psi_{13},~~
\psi^{14}=\tilde{B}\tilde{E}^{-1}\psi_{14},\nonumber\\
\psi^{23}&=&(\tilde{C}\tilde{D})^{-1}\psi_{23},~~\psi^{24}=(\tilde{C}\tilde{E})^{-1}\psi_{24},~~
\psi^{34}=(\tilde{D}\tilde{E})^{-1}\psi_{34}\nonumber
\end{eqnarray}
By using Eq.(\ref{2}), we obtain the following set of equations (for
simplicity, we assume $A_{0}\equiv A_{0}^{i}$ for all $i$)
\begin{eqnarray}
&&\tilde{B}[c_{0}(\partial_{1}S_{0})^{2}-c_{1}(\partial_{0}S_{0})(\partial_{1}S_{0})
-eA_{0}c_{1}(\partial_{1}S_{0})]+\tilde{C}^{-1}[c_{0}(\partial_{2}S_{0})^{2}\nonumber\\
&&-c_{2}(\partial_{0}S_{0})c_{0}(\partial_{2}S_{0})-eA_{0}c_{2}(\partial_{2}S_{0})]
+\tilde{D}^{-1}[C_{0}(\partial_{3}S_{0})^{2}-c_{3}(\partial_{3}S_{0})(\partial_{0}S_{0})\nonumber\\
&&-eA_{0}c_{3}(\partial_{3}S_{0})]+\tilde{E}^{-1}[c_{0}(\partial_{4}S_{0})^{2}
-c_{4}(\partial_{4}S_{0})(\partial_{0}S_{0})-eA_{0}c_{4}(\partial_{4}S_{0})]\nonumber\\
&&+m^{2}c_{0}=0,\label{r1}
\end{eqnarray}
\begin{eqnarray}
&&\tilde{A}^{-1}[c_{0}(\partial_{1}S_{0})(\partial_{0}S_{0})-c_{1}(\partial_{0}S_{0})^{2}
-eA_{0}c_{1}(\partial_{0}S_{0})]+\tilde{C}^{-1}[c_{1}(\partial_{2}S_{0})^{2}\nonumber\\
&&-c_{2}(\partial_{1}S_{0})(\partial_{2}S_{0})]+\tilde{D}^{-1}[c_{1}(\partial_{3}S_{0})^{2}
-c_{3}(\partial_{3}S_{0})(\partial_{1}S_{0})]+\tilde{E}^{-1}[c_{1}(\partial_{4}S_{0})^{2}\nonumber\\
&&-c_{4}(\partial_{1}S_{0})c_{0}(\partial_{4}S_{0})]+eA_{0}\tilde{A}^{-1}[c_{0}
(\partial_{1}S_{0})-c_{1}(\partial_{0}S_{0})]+m^{2}c_{1}=0,\\&&\tilde{A}^{-1}[c_{2}
(\partial_{0}S_{0})^{2}-c_{0}(\partial_{0}S_{0})(\partial_{2}S_{0})
+eA_{0}c_{2}(\partial_{0}S_{0})]-\tilde{B}[c_{2}(\partial_{1}S_{0})^{2}\nonumber\\
&&-c_{1}(\partial_{1}S_{0})(\partial_{2}S_{0})]+\tilde{D}^{-1}[c_{3}(\partial_{2}S_{0})
(\partial_{3}S_{0})-c_{2}(\partial_{3}S_{0})^{2}]+\tilde{E}^{-1}[c_{4}(\partial_{2}S_{0})
(\partial_{4}S_{0})\nonumber\\&&-c_{2}(\partial_{4}S_{0})^{2}]+eA_{0}\tilde{A}^{-1}[c_{2}
(\partial_{0}S_{0})-c_{0}(\partial_{2}S_{0})+eA_{0}c_{2}]-m^{2}c_{2}=0,\label{r2}\\
&&\tilde{A}^{-1}[c_{3}(\partial_{0}S_{0})^{2}-c_{0}(\partial_{0}S_{0})(\partial_{3}S_{0})
+eA_{0}c_{3}(\partial_{0}S_{0})]-\tilde{B}[c_{3}(\partial_{1}S_{0})^{2}\nonumber\\
&&-c_{1}(\partial_{1}S_{0})(\partial_{3}S_{0})]-\tilde{C}^{-1}[c_{3}
(\partial_{2}S_{0})^{2}-c_{2}(\partial_{2}S_{0})(\partial_{3}S_{0})]\nonumber\\
&&+\tilde{E}^{-1}[c_{4}(\partial_{4}S_{0})(\partial_{3}S_{0})-c_{3}(\partial_{4}S_{0})^{2}]
+eA_{0}\tilde{A}^{-1}[c_{3}(\partial_{0}S_{0})\nonumber\\
&&-c_{0}(\partial_{3}S_{0})+eA_{0}c_{3}]-m^{2}c_{3}=0,\label{r3}\\&&\tilde{A}^{-1}[c_{4}
(\partial_{0}S_{0})^{2}-c_{0}(\partial_{0}S_{0})(\partial_{4}S_{0})
+eA_{0}c_{4}(\partial_{0}S_{0})]-\tilde{B}[c_{4}(\partial_{1}S_{0})^{2}\nonumber\\
&&-c_{1}(\partial_{1}S_{0})(\partial_{4}S_{0})]-\tilde{C}^{-1}[c_{4}
(\partial_{2}S_{0})^{2}-c_{2}(\partial_{2}S_{0})(\partial_{4}S_{0})]\nonumber\\
&&-\tilde{D}^{-1}[c_{4}(\partial_{3}S_{0})^{2}-c_{3}(\partial_{3}S_{0})(\partial_{4}S_{0})]
+eA_{0}\tilde{A}^{-1}[c_{4}(\partial_{0}S_{0})\nonumber\\
&&-c_{0}(\partial_{4}S_{0})+eA_{0}c_{4}]-m^{2}c_{4}=0.\label{r4}
\end{eqnarray}
We carry out the separation of variables as
\begin{equation}
S_{0}=-(E-j{\check{\Omega}}_1)t+W(r)+\Theta(\zeta,\vartheta)+N\phi,
\end{equation}
where ${\check{\Omega}}_1$ is the angular velocity for BH given by Eq.(\ref{5D}).

For the above $S_{0}$ the preceding set of Eqs.(\ref{r1})-(\ref{r4})
can be written in terms of matrix equation
$\Lambda(c_{0},c_{1},c_{2},c_{3},c_{4})^{T}=0$, the elements of the
required matrix have the following form
\begin{eqnarray}
\Lambda_{00}&=&\tilde{B}\dot{W}^{2}+\tilde{C}^{-1}(\partial_{2}\Theta)^{2}+
\tilde{D}^{-1}(\partial_{3}\Theta)^{2}+\tilde{E}^{-1}N+m^{^{2}}\nonumber\\
\Lambda_{01}&=&\tilde{B}[(E-j{\check{\Omega}}_1)\dot{W}-eA_{0}\dot{W}],~~~~
\Lambda_{02}=\tilde{C}^{-1}(E-j{\check{\Omega}}_1)(\partial_{2}\Theta)\nonumber\\
\Lambda_{03}&=&\tilde{D}^{-1}(E-j{\check{\Omega}}_1)(\partial_{3}\Theta)
-\tilde{D}^{-1}eA_{0}(\partial_{3}\Theta),\nonumber\\
\Lambda_{04}&=&\tilde{E}^{-1}(E-j{\check{\Omega}}_1)N-\tilde{E}^{-1}jeA_{0},\nonumber\\
\Lambda_{10}&=&-\tilde{A}^{-1}(E-j{\check{\Omega}}_1)\dot{W}+eA_{0}\tilde{A}^{-1}\dot{W},\nonumber\\
\Lambda_{11}&=&-\tilde{A}^{-1}(E-j{\check{\Omega}}_1)^{2}+eA_{0}(E-j{\check{\Omega}}_1)\tilde{A}^{-1}
+\tilde{C}^{-1}(\partial_{2}\Theta)^{2}\nonumber\\
&+&\tilde{D}^{-1}(\partial_{3}\Theta)^{2}
+\tilde{E}^{-1}N^{2}+eA_{0}\tilde{A}^{-1}(E-j{\check{\Omega}}_1)+m^{2},\nonumber\\
\Lambda_{12}&=&-\tilde{C}^{-1}\dot{W}(\partial_{2}\Theta),~~~
\Lambda_{13}=-\tilde{D}^{-1}\dot{W}(\partial_{3}\Theta),\nonumber\\
\Lambda_{14}&=&-\tilde{E}^{-1}\dot{W}N,~~~\Lambda_{20}=\tilde{A}^{-1}(E-j{\check{\Omega}}_1)
(\partial_{2}\Theta)-\tilde{A}^{-1}eA_{0}(\partial_{2}\Theta),\nonumber\\
\Lambda_{21}&=&\tilde{B}\dot{W}(\partial_{2}\Theta),\nonumber
\end{eqnarray}
\begin{eqnarray}
\Lambda_{22}&=&\tilde{A}^{-1}(E-j{\check{\Omega}}_1)^{2}-eA_{0}(E-j{\check{\Omega}}_1)\tilde{A}^{-1}
-\tilde{B}\dot{W}^{2}\nonumber\\&-&\tilde{D}^{-1}(\partial_{3}\Theta)^{2}
-\tilde{E}^{-1}N^{2}+eA_{0}\tilde{A}^{-1}[eA_{0}-(E-j{\check{\Omega}}_1)]-m^{2},\nonumber\\
\Lambda_{23}&=&\tilde{D}^{-1}(\partial_{2}\Theta)(\partial_{3}\Theta),~~~
\Lambda_{24}=\tilde{E}^{-1}(\partial_{2}\Theta)N,\nonumber\\
\Lambda_{30}&=&\tilde{A}^{-1}(E-j{\check{\Omega}}_1)(\partial_{3}\Theta)-eA_{0}\tilde{A}^{-1}(\partial_{3}\Theta),~~~
\Lambda_{31}=\tilde{B}\dot{W}(\partial_{3}\Theta),\nonumber\\
\Lambda_{32}&=&\tilde{C}^{-1}(\partial_{2}\Theta)(\partial_{3}\Theta),\nonumber\\
\Lambda_{33}&=&\tilde{A}^{-1}(E-j{\check{\Omega}}_1)^{2}-e
A_{0}\tilde{A}^{-1}(E-j{\check{\Omega}}_1)
-\tilde{B}\dot{W}^{2}-\tilde{E}^{-1}N^{2}\nonumber\\&-&\tilde{C}^{-1}(\partial_{2}\Theta)^{2}-m^{2}-
eA_{0}\tilde{A}^{-1}[(E-j{\check{\Omega}}_1)-eA_{0}],\nonumber\\
\Lambda_{34}&=&\tilde{E}^{-1}j(\partial_{3}\Theta),~~~
\Lambda_{40}=\tilde{A}^{-1}((E-j{\check{\Omega}}_1)N-eA_{0}\tilde{A}^{-1}j,~~~
\Lambda_{41}=\tilde{B}\dot{W}N,\nonumber\\
\Lambda_{42}&=&\tilde{C}^{-1}(\partial_{2}\Theta)N,~~~
\Lambda_{43}=\tilde{D}^{-1}(\partial_{3}\Theta)N,\nonumber\\
\Lambda_{44}&=&\tilde{A}^{-1}(E-j{\check{\Omega}}_1)^{2}-eA_{0}\tilde{A}^{-1}(E-j{\check{\Omega}}_1)
-\tilde{B}\dot{W}^{2}-\tilde{C}^{-1}(\partial_{2}\Theta)^{2}\nonumber\\
&-&\tilde{D}^{-1}(\partial_{3}\Theta)^{2}-m^{2}-
eA_{0}\tilde{A}^{-1}[(E-j{\check{\Omega}}_1)-eA_{0}].\nonumber
\end{eqnarray}
For the non-trivial solution, the determinant $\Lambda$ is equal to
zero and using the same technique as discussed in the previous
section, we get
\begin{equation}
Im W^{\pm}=\pm \int\sqrt{\frac{(E-eA_{0}
-j{\check{\Omega}}_1)^{2}+\tilde{X}}{\tilde{A}\tilde{B}}}=\pm \iota \pi
\frac{(E-eA_0-j{\check{\Omega}}_1)}{2 \kappa(r_{+})}
\end{equation}
where
\begin{eqnarray}
\tilde{X}=-\tilde{A}\tilde{C}^{-1}(\partial_{2}\Theta)^{2}
-\tilde{A}\tilde{D}^{-1}(\partial_{3}\Theta)^{2}-\tilde{A}m^{2}-
\tilde{E}^{-1}(\partial_{2}\Theta)N.
\end{eqnarray}
Since BH given by Eq.(\ref{5D}) is non-rotating, so ${\check{\Omega}}_1=0.$
The surface gravity for this BH is given by \cite{R30}
\begin{equation}
{\kappa(r_{+})}=\frac{2r_{+}^{6}+r^{4}_{+}(1+\Sigma^{3}_{i=1}q_{i})-
\prod^{3}_{i=1}q_{i}}{r^{2}_{+}\sqrt{\prod^{3}_{i=1}(r^{2}_{+}+q_{i})}.}
\end{equation}
The required tunneling probability as discussed in the previous section is
\begin{equation}
\tilde{\Gamma}=\frac{\tilde{\Gamma}_{emission}}{\tilde{\Gamma}_{absorption}}
=e^{-4ImW^{+}}=e^{-2\pi
\frac{(E-eA_0)[{r^{2}_{+}\sqrt{\prod^{3}_{i=1}(r^{2}_{+}
+q_{i})}}]}{{2r_{+}^{6}+r^{4}_{+}(1+\sum^{3}_{i=1}q_{i})-
\prod^{3}_{i=1}q_{i}}}}\nonumber
\end{equation}
The Hawking temperature in this case is given by
\begin{equation}\label{abbas1}
\tilde{T}_{H}=\frac{[{2r_{+}^{6}+r^{4}_{+}(1+\sum^{3}_{i=1}q_{i})-
\prod^{3}_{i=1}q_{i}}]}{2\pi
{r^{2}_{+}\sqrt{\prod^{3}_{i=1}(r^{2}_{+}+q_{i})}}}.
\end{equation}
The Hawking temperature is related to energy $E$, potential $A_0$,
angular momentum $j$, the radial coordinate at the outer horizon
$r_{+}$ and charge $q_{i}$. We would like to mention that the
Hawking temperature of charged vector particles given by
Eq.(\ref{abbas1}) is same as the Hawking temperature of $5D$ gauged
super-gravity BH in Eq.(9) in Ref.\cite{R30}.

\section{Outlook}

During the tunneling process when a particle with electropositive
energy crosses the horizon, it seems as Hawking radiation. Likewise,
a particle with electronegative energy burrows inweave, it is
assimilated by the BH, so its mass falls and at last disappears.
Thus, the movement of the particles may be in the configuration of
outgoing and incoming, the carry out particle's action turns out
complex and real, respectively. The emission rate of a tunneling
particles from the BH is associated with the imaginary component of
the particles action, which infact is related to the Boltzmann
factor based on the Hawking temperature.

In this paper, we have extended the work of vector particles
tunneling for more generalized BHs in $4D$ and $5D$ spaces and
recovered their corresponding Hawking temperatures at which
particles tunnel through horizons. For this purpose, we have used
Proca equation with the background of electromagnetism to
investigate the tunneling of charged vector particles from
accelerating and rotating BHs $4D$ and $5D$ BHs having electric and
magnetic charges with a NUT parameter. We have implemented the WKB
approximation to Proca equation, which leads to the set of field
equations, then use separation of variables to solve these
equations. Solving for the radial part by using the determinant of
coefficient matrix equal to zero. Using surface gravity, we have
formulated the tunneling probability and Hawking temperature for
both BHs at the outer horizon. All these quantities depend on the
defining parameters of the BHs. It is worth while to mention here
that the back-reaction effects of the emitted particle on the BH
geometry and self-gravitating effects have been neglected, the
derived Hawking temperature only is a leading term. So that one does
not need to calculate the appropriate solution of the semi-classical
Einstein field equations for the geometry of background BH in
equilibrium with its Hawking radiation \cite{R36a}.

From our analysis we have concluded that, Hawking temperature at
which particles tunnel through the horizon is independent of the
species of particles. In particular \textit{nature of background BHs
geometries}, for the particles having different spins (either
spin-up or down) or zero spin, the tunneling probabilities will be
same by considering semi-classical effects. Thus, their
corresponding Hawking temperatures must be same for all kinds of
particles. For both cases, we have carried out the calculations for
more general BHs, i.e., a pair of charged accelerating and rotating
BHs with NUT parameter (which is more general BHs as compared to BH
taken in \cite{A36}) and a BH in $5D$ gauged super-gravity. Our
findings are similar with the statement i.e., temperature of
tunneling particles is independent of species of the particles, this
result is also valid for different coordinate frames by using
specific coordinate transformations. The authors of Ref.\cite{A36}
have been proved for Kerr BH (only rotating), while we have proved
for more generalized BHs. Hence, the conclusion still holds if
background BHs geometries are more generalized.

\end{document}